\documentclass[12pt]{article}
\usepackage{amssymb}
\newcommand{\calle}[1]{(\ref{#1})}
\newcommand{\beq}{ \begin{equation} }
\newcommand{\eeq}{ \end{equation} }
\newcommand{\beqn}{ \begin{eqnarray} }
\newcommand{\eeqn}{ \end{eqnarray} }
\newcommand{\bb}{ \begin{eqnarray*} }
\newcommand{\ee}{ \end{eqnarray*} }

\begin{document}

\begin{center}
{\bf {\Large The Conformal Properties of
Liouville Field Theory on $\mathbb{Z}_N$-Riemann Surfaces}}

\vspace{1cm}

        S.A. Apikyan $^\dagger$ \\
        Theoretical Physics Departament\\
        Yerevan Physics Institute\\
        Alikhanyan Br.st. 2, Yerevan, 375036 Armenia\\
       
        \vspace{3mm}
        C.J. Efthimiou$^\ddagger$ \\
        Newman Laboratory of Nuclear Studies\\
        Cornell University\\ 
        Ithaca, NY 14853, USA
\end{center}

\vspace{1cm}

\begin{abstract}
The Liouville field theory on $\mathbb{Z}_N$-Riemann surfaces 
is studied and it is shown that it  decomposes into a Liouville
field theory on the sphere and $N-1$ free boson theories.
Also, the partition function of the Liouville
field theory on the $\mathbb{Z}_N$-Riemann surfaces 
is expressed as a product of the correlation function for the
Liouville vertex operators on the sphere and a number of twisted fields.
\end{abstract}

\vfill
\hrule
\ \\
\noindent
$\dagger$ e-mail: apikyan@lx2.yerphi.am\\
$\ddagger$ e-mail: costas@mail.lns.cornell.edu

\newpage
\indent

\section{Introduction}

The main motivation to study two-dimensional
Liouville field theory (LFT) is its relation to the string theories
and the realization that it provides an effective theory of 2D quantum
gravity.
However, despite significant progress in understanding
the classical Liouville theory, 
our understanding of the quantum Liouville field theory 
is quite limited. 
However, since the 80's, essential  
progress has been achieved in the understanding of
quantum LFT (\cite{AG,Sei} and references therein).
The interest on LTM was intensified with the development
of the matrix model approach that confirmed results
obtained with the LFT (\cite{GM} and references therein).
Recently, the interest on LFT has again been renewed
since an analytic expression for the three-point correlation
function of the Liouville vertex operators  has been
consrtucted \cite{DO1,DO2,ZZ}. 

The present work is organized as follows. In the first part of 
section \ref{sec:2}, a brief description of the LFT
on $\mathbb{Z}_N$-surfaces is given. In the second part of section  
\ref{sec:2},
we use Polyakov's proposal, to express the partition function
of the LFT on a $Z_N$-surface  as a partition function of a LFT 
on a sphere and free scalar
field theories with inserted Liouville vertex 
operators and twisted fields. 
The present paper  generalizes a  previous work 
of one of the authors \cite{A}.

\section{LFT on $\mathbb{Z}_N$-Riemann surfaces.}
\label{sec:2}

A $\mathbb{Z}_N$-symmetric Riemann surface $X_g^{(N)}$ 
of genus $g\geq1$ is determined by an 
algebraic equation of the form
$$
  y^N(z)=\prod_{i=1}^{h}(z-\omega_{i})^{n_i}~, \qquad n_i>1~,
$$
i.e. $X_g^{(N)}$-is an $N$-sheeted covering of a Riemann sphere.
The genus $g$ of a $\mathbb{Z}_N$-Riemann surface
$$
   g=\frac{(N-1)(h-2)}{2}
$$
can be calculated using the Riemann-Hurwitz theorem.
The algebraic equation of the  $\mathbb{Z}_N$-surface 
has $h$ complex parameters;
therefore the moduli space $\mathcal{M}_{\mathbb{Z}_N}$
has dimension
$$
 \dim\mathcal{M}_{\mathbb{Z}_N}=h-3=\frac{2g}{N-1}-1~.
$$
Comparing with the dimension of the moduli space
$\mathcal{M}$ for generic Riemann surfaces,
$$
   \dim\mathcal{M}=\cases{1, & if $~g=1~,$\cr
                         3g-3, & if $~g>1~,$\cr}
$$
we can conclude that the $\mathbb{Z}_N$-surfaces do not
contain all Riemann surfaces.

We start with Liouville theory in the conformal gauge.
The action is given by
$$
  \mathcal{S}=\int_{\mathbb{Z}_N}\, d^2y\,
  \left\lbrack\frac{1}{4\pi}(\partial_a \phi)^2 +
  \mu \, e^{2b\phi} + \frac{Q}{4\pi}\, \hat{R}\phi\right\rbrack ~,
$$
where $b$ and $\mu$ are the coupling and cosmological 
constants respectively. We have fixed a fiducial metric
$\hat{g}$ on a given  surface 
with curvature $\hat{R}$ normalized by
$$
   \frac{1}{4\pi}\int_{\mathbb{Z}_N}\,d^2y\, \sqrt{\hat{g}}\hat{R}=2(1-g)
    ~.
$$

We label the $N$ sheets of the Riemann $\mathbb{Z}_N$-surface
$X_g^{(N)}$ by the numbers $l=0,1,...,N-1$:
\begin{equation}
\label{eq:7}
  y^{(l)}(z)=\omega^{l}\prod_{j=1}^{h}(z-\omega_j)^{n_j/N}
  ~.
\end{equation}
Under the map \calle{eq:7}, the Lagrangian density $\mathcal{L}(\phi(y))$,
the energy-momentum tensor $T(\phi(y))$, and the Liouville fields 
$\phi(y)$
on the $\mathbb{Z}_N$-Riemann surface map into branches:
$\mathcal{L}^{(l)}(\phi^{(l)}(z)), T^{(l)}(z)$, and $\phi^{(l)}(z)$, 
$l=0,1,...,N-1$ respectively.

Let $\Omega(z)$ be the conformal factor of the metric under
conformal transformations of the coordinates.
The Liouville branch fields $\phi^{(l)}(z)$
transform like a logarithm of the conformal
factor:
$$
   \phi^{(l)}(\omega,\bar{\omega})=\phi^{(l)}(z,\bar{z})-
   \frac{Q}{2}\log\mid\Omega'(z)\mid^2~,
$$
where
$$
   Q=b+\frac{1}{b}~.
$$
On each sheet, we have a holomorphic Liouville
energy-momentum tensors
$$
     T^{(l)}(z)=-(\partial \phi^{(l)}(z,\bar{z}))^2 + 
     Q\, \partial^2\phi^{(l)}(z,\bar{z})~,
$$
with the Liouville central charge
$$
\hat{c}=1+6Q^2 ~.
$$

As usual, we assume that the fields $(T^{(l)}(z),\phi^{(l)}(z,\bar{z}))$
on the $N$-sheeted covering may be considered as vector fields
on $\mathbb{C}P^1$. When the argument of these vector fields encircles
the branch points, they transform among themselves according
to a certain monodromy matrix.
This monodromy matrix forms a representation 
of the first homotopy group $\pi_1(\mathbb{C}P^1/\cup \omega_j)$
which in our case  is just $\mathbb{Z}_N$.
It is convenient to pass to a basis \cite{AE} 
in which the generators of monodromy group are diagonal:
$$
\begin{array}{l}
   \phi_{(k)}(z,\bar{z})=\sum\limits_{l=0}^{N-1}
   \omega^{-kl} \, \phi^{(l)}(z,\bar{z})~, \\   \\
   T_{(k)}(z)=\sum\limits_{l=0}^{N-1}
   \omega^{-kl}\, T^{(l)}(z)~.
\end{array}
$$
The ``bosonization rule" for the operators $T_{(k)}$
in the diagonal basis can be written as follows:
$$
   T_{(k)}=
   -\frac{1}{N}\sum_{s=0}^{N-1}\partial \phi_{(s)}
   \partial \phi_{(k-s)} + 
   Q \partial^2 \phi_{(k)}~.
$$
In particular,  the form of the Liouville energy-momentum 
tensor is given by
\beq
\label{eq:14}
   T_{(0)}=
   -\frac{1}{N}\partial \phi_{(0)}\partial \phi_{(0)} + 
   Q \partial^2 \phi_{(0)}-
   \frac{1}{N}\sum_{s\neq0}^{N-1}\partial \phi_{(s)}
   \partial \phi_{(-s)}  
\eeq
and the corresponding Liouville central charge is
$$
    c=N\, (1+6Q^2)~.
$$

According to the previous results,
the original theory splits into a sum of a 
LFT on a sphere with the central charge $c_s=1+6Q^2N$
and $N-1$ free field theories $\phi_{(s)}$, $s=1,2,...,N-1$
with central charge $c_f=1$.

Under a holomorphic transformation of the coordinates,
the Liouville field $\phi_{(0)}\equiv \Phi$
and the free fields $\phi_{(k)}$ $(k\neq 0)$
transform as follows:
$$
\begin{array}{l}
   \Phi(\omega,\bar{\omega})=\Phi(z,\bar{z})-
   \frac{N}{2}Q\log\mid\Omega'(z)\mid^2~,  \\
   \\
   \phi_{(k)}(\omega,\bar{\omega})=\phi_{(k)}(z,\bar{z})~,
    \qquad k\neq 0~.
\end{array}
$$
We rewrite \calle{eq:14} in terms of new variables:
$$
   T=-\frac{1}{N}\partial \Phi\partial \Phi + 
   Q \partial^2 \Phi-
   \frac{1}{N}\sum_{s\neq0}^{N-1}\partial \phi_{(s)}
   \partial \phi_{(-s)}~.  
$$
According to the monodromy properties \cite{AE}
of the vector fields on $\mathbb{C}P^1$, we have to define two
kinds of ``Liouville vertex operators". The first kind
is ``untwisted vertex operators": 
$$
   V_{[0]}(z,\bar{z})=
   e^{2\alpha\Phi(z,\bar{z})}
   e^{i\sum_{s\neq 0}\alpha_{(s)}\phi_{(s)}(z,\bar{z})}
   ~.
$$
with dimensions
\beq
\label{eq:19}
   \Delta_{[0]}=
    2\alpha(Q-\alpha)+\frac{N}{2}\sum_{s\neq0}\alpha_{(s)}\alpha_{(-s)}
   ~.
\eeq
The physical LFT space of states is defined by the following charge
\cite{CT,GN}:
$$
     \alpha=ip+\frac{Q}{2}~.
$$
Substituting  this value in the expression \calle{eq:19}, we find  
$$
   \Delta_{[0]}=\frac{Q^2}{2}+2p^2+\frac{N}{2}\sum_{s\neq0}
    \alpha_{(s)}\alpha_{(-s)}~.
$$
The second kind of vertex operators is the ``twisted vertex operators"
which have the form:
$$
   V_{[k]}(z,\bar{z})=
   e^{2\gamma \Phi(z,\bar{z})}\sigma_{k}(z|1)\sigma_{k}(z|2)
   \dots\sigma_{k}(z|N-1)~.
$$
In the above formula,
$\sigma_{k}(z|l)$ is a twist fields  having 
 dimension
$$
   \Delta_{kl}=\frac{1}{4}[\{\frac{kl}{N}\}-\{\frac{kl}{N}\}^2]
   ~,
$$
where the symbol $\{x\}$ denotes the fractional part of $x$.
Thus, the twisted vertex operators have dimensions:
$$
  \Delta_{[k]}=
  2\gamma(Q-\gamma)+\frac{N^2-1}{24N}~.
$$

We  now proceed to construct the partition function of 
the LFT on the $\mathbb{Z}_N$-surface $X_g^{(N)}$
by making  use of the above results.
According to a main proposal of Polyakov \cite{T},
a ``summation" over a smooth metric with  insertion
of vertex operators at some points  should be equivalent to the 
``summation" over a metric with singularities at the insertion
points and without insertion of any vertex operators.
Therefore, the partition function of the Liouville field theory 
on the $\mathbb{Z}_N$-surface,
$$
  Z_g=\Bigg\lmoustache D\phi \,
  \exp \left\lbrack -\int \frac{1}{4\pi}(\partial_a \phi)^2 +
  \mu\, e^{2b\phi} + \frac{Q}{4\pi}\hat{R}^g\phi \right\rbrack 
  ~,
$$
can be represented by the expression
\beqn
  Z_g &=& \Bigg\lmoustache D\Phi\,
  \exp\left\lbrack-\int \frac{1}{4\pi N}(\partial_a \Phi)^2 +
  \mu e^{2b\Phi} + \frac{Q}{4\pi}\hat{R}^{g=0}\Phi\right\rbrack
  \nonumber\\
   &\times& \prod_{s\neq0} \,D \phi_{(s)}\, 
   \exp\left\lbrack-\int\frac{1}{4\pi N}
   \sum_{s\neq0}\partial\phi_{(s)}\partial\phi_{(-s)}\right\rbrack
   \nonumber\\
   &\times& \prod_{i=1}^{h}e^{2\gamma_i \Phi(\omega_i,\bar{\omega}_i)}
   \, \sigma_{k_i}(\omega_i|1)\sigma_{k_i}(\omega_i|2)
   ...\sigma_{k_i}(\omega_i|N-1)~,
\label{eq:28}
\eeqn
where 
$$
b=\frac{Q}{2}\pm\sqrt{\frac{Q^2}{4}-\frac{1}{2}}~.
$$
In order to evaluate the partition function written above,
we will first integrate over the zero mode of $\Phi$.
After this integration is performed, we find
\beqn
   Z_g &=& (-\mu)^s\frac{\Gamma(-s)}{b}
   \nonumber\\
 &\times&  \Bigg\lmoustache\, D\tilde{\Phi}
 \, \exp\left\lbrack -\int \frac{1}{4\pi N}(\partial\tilde{\Phi})^2 +
 \frac{Q}{4\pi}\hat{R}^{g=0}\tilde{\Phi}\right\rbrack
 \, \left(\int e^{2b\tilde{\Phi}}\right)^s
  \, \prod_{i=1}^{h}e^{2\gamma_i\tilde{\Phi}(\omega_i,\bar{\omega}_i)}
  \nonumber\\
 &\times&  \Bigg\lmoustache \, \prod_{s\neq0}
 D \phi_{(s)}\, \exp\left\lbrack -\int\frac{1}{4\pi N}
\sum_{s\neq0}\partial\phi_{(s)}\partial\phi_{(-s)}\right\rbrack
\nonumber\\
&\times& \prod_{i=1}^{h}
\sigma_{k_i}(\omega_i|1)\sigma_{k_i}(\omega_i|2)
...\sigma_{k_i}(\omega_i|N-1) ~,
\label{eq:30}
\eeqn
where 
\beq
\label{eq:31}
  \sum_{i=1}^h\gamma_i=Q-sb~,
\eeq
and $\tilde\Phi$ denotes  fields orthogonal to the zero mode \cite{GL}.
The correlation function in \calle{eq:28}
of the fields $\sigma_k$ is determined  by:
$$
 \langle\prod_{i=1}^{h}
\sigma_{k_i}(\omega_i|1)\sigma_{k_i}(\omega_i|2)
...\sigma_{k_i}(\omega_i|N-1)\rangle =\prod\mid\omega_i-\omega_j\mid^
{-2\gamma_{ij}}({\rm det}\widehat{W})^{-1/2}~,
$$
where the  matrix $\widehat{W}$ is the period matrix
and $\gamma_{ij}$
forms a basis in $H_1(X_g^{(N)},\mathbb{Z})$ \cite{BR}.
Moreover, it is well-known that the multi-point correlator 
\calle{eq:30}
leads to the partition function of the free scalar fields 
under the Ramond boundary condition \cite{K}.
The first correlation function in \calle{eq:28} is not a free 
field correlator because, in general, the power $s$
is not a positive integer. However, for integer values, 
$s=n\in\mathbb{Z}$, the partition
function $Z_g$ exhibits a pole in the $\sum\gamma_i$
with the residue being equal to the corresponding
perturbative integral
$$
  \mathrel{\mathop {\rm Res}\limits_{\sum\gamma_i=Q-nb}} 
  Z_g(\omega_1,...,\omega_h) =
 G^{(n)}(\omega_1,...,\omega_h) \Bigg|_
 {\sum\gamma_i=Q-nb}
  ~,
$$
where $G^{(n)}$ is the free field correlator
\bb
G^{(n)} &=& \frac{(-\mu)^n}{n!} \,
\Bigg\lmoustache\, D\Phi \, e^{-\int\frac{1}{4\pi N}
(\partial \Phi)^2 + Q \hat{R}^{g=0} \Phi}\\
&\times& \prod_{j=1}^{n}\int \sqrt{\hat{g}}\, e^{2b\Phi(x_j)}d^2x_j
 \, \prod_{i=1}^{h}e^{2\gamma_i \Phi(\omega_i,\bar{\omega}_i)}\\
&\times&
\Bigg\lmoustache\, \prod_{s\neq0}\, D\phi_{(s)}\,
\exp\left\lbrack -\frac{1}{4\pi N}\int\sum_{s\neq0}
\partial\phi_{(s)}\partial\phi_{(-s)}\right\rbrack\\
&\times& \prod_{i=1}^{h}\,
\sigma_{k_i}(\omega_i|1)\sigma_{k_i}(\omega_i|2)
...\sigma_{k_i}(\omega_i|N-1)~.
\ee
This is just the  $n$-th term in the naive perturbation  of $Z_g$ 
in powers of $\mu$.
So, the LFT partition function on the $\mathbb{Z}_N$-surface 
has been
reduced to the Liouville correlation function on the sphere
with inserted Liouville vertex operators (with charges $\gamma_i$)
and to a correlation function of the twisted fields \cite{Z}.
The residue of the LFT partition function on the $\mathbb{Z}_N$-surface
at the poles are the correlation functions of the free field
theories on the $\mathbb{Z}_N$-surface.

Let us consider the special case of the Liouville field 
theory on an elliptic curve, i.e. $N=2,h=4$. We can rewrite
the partition function $Z_1$ for this particular case as:
\bb
 Z_1 &=& \Bigg\lmoustache\, D\Phi\,
 \exp\left\lbrack -\int \frac{1}{8\pi N}(\partial_a \Phi)^2 +
 \mu e^{2b\Phi} + \frac{Q}{4\pi}\hat{R}^{g=0}\Phi\right\rbrack\\
 &\times& \Bigg\lmoustache\, D\phi\, \exp\left\lbrack -\int\frac{1}{8\pi}
 \partial\phi\partial\phi\right\rbrack\,
 \prod_{i=1}^{4}e^{2\gamma_i \Phi(\omega_i,\bar{\omega}_i)}
 \sigma_{\epsilon_i}(\omega_i,\bar{\omega}_i)
  ~.
\ee
The residue of $Z_1$ (corresponding to \calle{eq:31})
will be equal to the conformal four-point function
$$
   \mathrel{\mathop {\rm Res}\limits_{\sum\gamma_i=Q-nb}} 
   Z_1(\omega_1,...,\omega_4) =
   \left.G^{(n)}(\omega_1,...,\omega_4)\right|_
   {\sum\gamma_i=Q-nb}~,
$$
where $G^{(n)}$ has the form
\bb
  &&  G_{\gamma_1\gamma_2\gamma_3\gamma_4}^{(n)}
    (\omega_1,\omega_2,\omega_3,\omega_4)=
    \frac{\left(-\mu \right)^n}{n!}
\, \Bigg\lmoustache\, D\Phi\,  e^{-\int\frac{1}{8\pi}
(\partial\Phi)^2 + Q\hat{R}^{g=0}\Phi}\\
&&  \times\prod_{j=1}^{n}\int e^{2b\Phi(x_j)}d^2x_j
\prod_{i=1}^{4}e^{2\gamma_i\Phi(\omega_i,\bar{\omega}_i)}
\Bigg\lmoustache_{\phi \in S^1} D\phi\, 
\exp\left\lbrack -\frac{1}{8\pi}
\int \partial\phi\partial\phi \right\rbrack~.
\ee
We can thus conclude that the Liouville partition function on
an elliptic curve reduces to the four-point correlation function
of the Liouville vertex operators on the sphere and the partition function
of the free field theory where integration goes over the compactified 
fields with  Ramond boundary conditions.


\begin{thebibliography}{99}

\bibitem{AG}
  L. Alvarez-Gaum\'e and C. Gomez, in ``String Theory and Quantum Gravity",
  J Harvey,  R. Iengo, K.S. Narain, S. Randjbar-Daemi, and H.
  Verlinde, eds, World Scientific, 1992.
\bibitem{Sei}
  N. Seiberg, in ``Reandom Surfaces and Quantum Gravity",
  O. Arvarez, E. Marinari, and P. Windey, eds, Plenum Press 1990.
\bibitem{GM}
  P. Ginsparg and G. Moore, in ``Recent Directions in Particle Theory:
  From Superstrings and Black Holes to the Standard Model", 
  J. Harvey and J. Polchinski, eds, World Scientific 1993.
\bibitem{DO1}
   H. Dorn and H. Otto, Phys. Lett. {\bf B291} (1992) 39.
\bibitem{DO2} 
   H. Dorn and H. Otto, Nucl. Phys. {\bf B429} (1994) 375.
\bibitem{ZZ} 
   A.B. Zamolodchikov and Al.B. Zamolodchikov, hep-th/9506136.
\bibitem{GL}
   M. Goulian and M. Li, Phys. Rev. Lett. {\bf 66} (1991) 2051.
\bibitem{T} 
   L. Takhtajan, 
   in ``New Symmetry Principles in Quantum
   Field Theory", J. Frolich et al., eds,  Plenum Press 1992.
\bibitem{AE} 
   S.A. Apikyan and C.J. Efthimiou, hep-th/9610051.
\bibitem{CT} 
   T. Curtright and C. Thorn, Phys. Rev. Lett. {\bf 48} (1982) 1309;
   E. Braaten, T.Curtright, and C.Thorn, Phys. Lett. {\bf B118} (1982) 115;
   Ann. Phys. {\bf 147} (1983) 365.
\bibitem{GN} 
   J. Gervais and A. Neveu, Nucl. Phys. {\bf B238} (1984) 125;
   {\bf B238} (1984) 396; {\bf B257} [FS14] (1985) 59.
\bibitem{BR}
   M. Bershadsky and A. Radul, Int. Mod. Phys. {\bf A2} (1987) 165.
\bibitem{K} 
   V. Knizhnik, Commun. Math. Phys. {\bf 112} (1987) 567.
\bibitem{Z} 
   Al. B. Zamolodchikov, Nucl. Phys. {\bf B285} [FS19] (1987) 481;
   L. Dixon, D. Friedan, E. Martinec, and S. Shenker, Nucl. Phys. 
   {\bf B282} (1987) 13;
   C. Crnkovic, G. Sotkov, and M. Stanishkov,  Phys. Lett. {\bf B220}
   (1989) 397;
   S.A. Apikyan and C.J. Efthimiou, Phys. Lett. {\bf B359} (1995) 313;
   S.A. Apikyan and Al.B. Zamolodchikov,  ZhETF {\bf 92} (1987) 34.
\bibitem{A} 
   S.A. Apikyan, Phys. Lett. {\bf B388} (1996) 557.
\end{thebibliography}
\end{document}